\pdfoutput=1
\documentclass[twocolumn,floatfix,nofootinbib]{revtex4-1}

\usepackage[utf8]{inputenc}
\usepackage[T1]{fontenc}
\usepackage{lmodern} 
\usepackage[ngerman,english]{babel}
\usepackage{graphicx} 
\usepackage{microtype} 
\usepackage{mathtools} 
\usepackage{bbold} 

\usepackage{xcolor}

\usepackage{hyperref}
\hypersetup{colorlinks=true,urlcolor=blue,linkcolor=red,citecolor=green!60!black}

\usepackage{tikz}
\usetikzlibrary{arrows,shapes.misc,decorations.markings,decorations.pathmorphing,positioning,intersections}
\tikzset{cross/.style={cross out, draw=black, minimum size=2*(#1-\pgflinewidth), inner sep=0pt, outer sep=0pt},
cross/.default={1.5mm}}
\tikzset{mydash/.style={dashed, dash pattern=on 4pt off 5pt}}
\tikzset{
  vertex/.style={draw,shape=circle,fill=black,minimum size=3pt,inner sep=0pt},
  cross/.style={cross out, draw=black,thick, minimum size=6pt, inner sep=0pt, outer sep=0pt},
  external/.style={inner sep=2pt},
  plabel/.style={inner sep=2pt},
  blob/.style={circle,fill=black!20,minimum size=0.7cm,draw,thick},
  whiteblob/.style={circle,fill=white,minimum size=1.0cm,draw,thick},
  effective/.style={rectangle,fill=black!20,minimum size=0.5cm,draw,thick},
  vev/.style={shape=vev,draw,inner sep=2pt,thick},
  mass/.style={shape=cross,draw,thick},
  rscalar/.style={dashed,thick},
  mfermion/.style={thick},
  scalar/.style={postaction={decorate}, decoration={markings,mark=at position .55 with {\arrow{latex}}},dashed,thick},
  ooscalar/.style={postaction={decorate}, decoration={markings,mark=at position .7 with {\arrow{latex}}},dashed,thick},
  fermion/.style={postaction={decorate}, decoration={markings,mark=at position .55 with {\arrow{latex}}},thick},
  majfermion/.style={postaction={decorate}, decoration={markings,mark=at position .7 with {\arrow{latex}}},thick},
  oofermion/.style={postaction={decorate}, decoration={markings,mark=at position .85 with {\arrow{latex}}, mark=at position .35 with {\arrowreversed{latex}}},thick},
  iifermion/.style={postaction={decorate}, decoration={markings,mark=at position .35 with {\arrowreversed{latex}}, mark=at position .85 with {\arrow{latex}}},thick},
  gaugeboson/.style={decorate, decoration={snake},thick},
  gluon/.style={decorate, decoration={coil,amplitude=4pt, segment length=5pt},thick},
  photon/.style={decorate, decoration={snake},thick},
  dashdot/.style={dash pattern=on .4pt off 3pt on 4pt off 3pt,thick}}

\newcommand{\GeV}{\,\mathrm{GeV}}
\newcommand{\TeV}{\,\mathrm{TeV}}
\renewcommand{\Im}[1]{\,\mathrm{Im}\!\left[#1\right]}

\newcommand*{\rom}[1]{\text{\expandafter \MakeUppercase{\romannumeral #1}}}

\begin{document}


\title{Low-scale leptogenesis assisted by a real scalar singlet}

\def\andname{\hspace{-1ex}}
\author{Tommi Alanne}\email{tommi.alanne@mpi-hd.mpg.de}
\author{Thomas Hugle}\email{thomas.hugle@mpi-hd.mpg.de}
\author{Moritz Platscher,}\email{moritz.platscher@mpi-hd.mpg.de}

\affiliation{Max-Planck-Institut f\"ur Kernphysik,
Saupfercheckweg 1, 69117 Heidelberg, Germany}

\author{and Kai Schmitz}\email{kai.schmitz@pd.infn.it}

\affiliation{Dipartimento di Fisica e Astronomia, Universit\`a degli Studi di Padova,
Via Marzolo 8, 35131 Padova, Italy}

\affiliation{Istituto Nazionale di Fisica Nucleare (INFN), Sezione di Padova,
Via Marzolo 8, 35131 Padova, Italy}


\begin{abstract}
\noindent
Standard thermal leptogenesis in the type-I seesaw model 
requires very heavy right-handed neutrinos (RHNs).
This makes it hard to probe this scenario experimentally and
results in large radiative corrections to the Higgs boson mass.
In this paper, we demonstrate that the situation is considerably different
in models that extend the Higgs sector by a real scalar singlet.
Based on effective-theory arguments, the extra scalar is always
allowed to couple to the heavy neutrinos via singlet Yukawa terms.
This opens up new RHN decay
channels leading to larger $CP$ violation as well as to a stronger
departure from thermal equilibrium during leptogenesis.
As a consequence, the baryon asymmetry can be generated
for a lightest RHN mass as low as $500\GeV$ and without the need for
a highly degenerate RHN mass spectrum.
In fact, the requirement of successful leptogenesis via the Higgs portal
coupling singles out an interesting parameter region that
can be probed in on-going and future experiments.
We derive a semianalytical fit function for the final baryon asymmetry
that allows for an efficient study of parameter space, thus enabling us
to identify viable parameter regions.
Our results are applicable to a wide range of models
featuring an additional real scalar singlet.
\end{abstract}


\maketitle


\section{Introduction}
\label{sec:intro}


The origin of the baryon asymmetry of the Universe (BAU), typically
quantified in terms of the cosmic baryon-to-photon ratio
$\eta_B^{\rm obs} \simeq 6.1 \times 10^{-10}$~\cite{Aghanim:2018eyx,Tanabashi:2018oca}, is
one of the greatest mysteries in fundamental physics.
An attractive approach towards this problem consists in baryogenesis
via leptogenesis~\cite{Fukugita:1986hr}  in the context of the type-I seesaw
mechanism~\cite{Minkowski:1977sc,Yanagida:1979as,Yanagida:1980xy,GellMann:1980vs,Mohapatra:1979ia}.
Leptogenesis is motivated by the fact that it links the generation
of the BAU at high energies to the phenomenology of neutrino
oscillations at low energies.
Depending on further model assumptions, this connection allows one to derive
testable predictions in the neutrino sector based on the requirement of successful
leptogenesis (see Refs.~\cite{Dev:2017trv,Drewes:2017zyw,Dev:2017wwc,Biondini:2017rpb,
Chun:2017spz,Hagedorn:2017wjy} for a recent series of review articles).


The type-I seesaw model postulates the existence of at least two
right-handed neutrinos (RHNs), $N_i$, that transform as complete singlets
under the Standard Model (SM) gauge group and whose Majorana
masses, $M_i$, violate lepton number by two units
(see, e.g., Refs.~\cite{Bambhaniya:2016rbb,Rink:2016vvl,Rink:2016knw} for minimal
realizations of the type-I seesaw model).
The $CP$-violating out-of-equilibrium decays of these neutrinos
in the early Universe generate a primordial lepton asymmetry
that is subsequently converted into a primordial baryon asymmetry
by means of electroweak (EW) sphaleron processes~\cite{Kuzmin:1985mm}.
However, in the standard scenario of thermal leptogenesis, the amount of $CP$
violation in RHN decays turns out to be suppressed by the tiny masses of
the active SM neutrinos, which results in a strong lower bound
on the RHN mass scale~\cite{Davidson:2002qv}.
A simplified treatment, not taking into account the dynamics in flavor space,
leads to the conclusion that the lightest RHN mass, $M_1$, must be at least
of $\mathcal{O}\left(10^9\right)\GeV$~\cite{Buchmuller:2002rq,Giudice:2003jh,Buchmuller:2004nz}.
The inclusion of flavor effects, on the other hand, can lower this bound by up to three
orders of magnitude, $M_1 \gtrsim 10^6\GeV$~\cite{Hambye:2003rt,Blanchet:2008pw,Antusch:2009gn,Moffat:2018wke}\,---\,but
certainly not to the extent that the RHN mass scale would approach
the energy range that is probed by current or near-future colliders.
The direct observation of heavy Majorana neutrinos in terrestrial
experiments therefore seems to be out of reach in the standard scenario,
at least in the foreseeable future.
In addition to this phenomenological drawback, standard thermal
leptogenesis is also plagued by a pressing theoretical issue:
large radiative corrections that arise
in the RHN sector at the one-loop level.
Indeed, the large hierarchy between the RHN mass scale and the EW
scale, $M_i \gg v_{\rm ew} \simeq 246\GeV$, threatens to destabilize the Higgs mass,
which necessitates some severe fine-tuning of input parameters~\cite{Vissani:1997ys},
as long as no other stabilization mechanism is at work.%
\footnote{Note, however, that the radiative corrections in the type-I seesaw
model may also be used to explain the very origin of the SM Higgs potential.
This is the basic idea behind the so-called
``neutrino option''~\cite{Brivio:2017dfq,Brdar:2018vjq,Brivio:2018rzm,Brdar:2018num},
which assumes that the classical SM Lagrangian satisfies scale-invariant
boundary conditions at high energies.}


Together, these observations motivate efforts to seek alternative
realizations of the leptogenesis paradigm that 
allow for the possibility
of successful baryogenesis at a significantly lower RHN mass scale.
Two popular approaches in this direction are
(i) resonant leptogenesis~\cite{Pilaftsis:1997jf,Pilaftsis:2003gt},
which assumes a highly degenerate RHN mass spectrum, and
(ii) leptogenesis via the Akhmedov-Rubakov-Smirnov mechanism
of RHN oscillations~\cite{Akhmedov:1998qx}
(see Refs.~\cite{Abada:2018oly,Ghiglieri:2018wbs} for very recent work
based on these ideas).
Alternatively, one may consider slight modifications
of the type-I seesaw model that provide one with a larger
parametric freedom in generating the BAU~\cite{AristizabalSierra:2007ur}.
One such example is, e.g., leptogenesis in  
the scotogenic
model of radiative neutrino masses~\cite{Ma:2006km}.
This model promotes the type-I seesaw sector to a ``dark sector'',
where the sterile neutrinos couple to a second, dark Higgs doublet,
and in which all fields transform odd under a new discrete $\mathbb{Z}_2$
symmetry.
As recently shown, 
the additional parameters present
in this scenario allow one to lower the energy scale of leptogenesis
down to RHN masses of $\mathcal{O}\left(10\right)\TeV$~\cite{Hugle:2018qbw}.
For further studies of baryogenesis in the scotogenic model,
see Refs.~\cite{Borah:2018uci,Baumholzer:2018sfb,Huang:2018vcr,Borah:2018rca,Borah:2018enf}.


In this paper, we will 
study another opportunity for low-scale
leptogenesis that builds upon the general type-I seesaw framework.
As shown in Refs.~\cite{Sierra:2014sta,Dall:2014nma}, 
this mechanism, while being more minimal
than leptogenesis in the scotogenic model, allows for a significantly lower
RHN mass scale and, as we will demonstrate, 
manages to successfully generate the BAU for RHN masses 
even below the TeV scale and still does not require a strong RHN mass degeneracy.
While the scotogenic model assumes the existence of an extra scalar
doublet, we will consider in this paper the mere addition of an
extra real scalar singlet, $S$.
The scalar sector 
thus represents the simplest
conceivable extension of the SM Higgs sector.
In the main part of our analysis, we will demonstrate how the presence
of the scalar $S$ modifies some of the key quantities in the RHN sector
and develop a numerical as well as analytical understanding of the
ensuing leptogenesis scenario.
This analysis will enable us to identify viable parameter regions
that can be probed in on-going and future experiments.


The existence of an extra real scalar singlet is well motivated
and frequently encountered in various more
fundamental models.
If the scalar $S$ is stable, it represents, e.g., a viable and minimal
particle candidate for dark matter~\cite{Silveira:1985rk,McDonald:1993ex,Burgess:2000yq,Athron:2017kgt}.
Alternatively, it is possible to identify $S$ with the inflaton field,
which is responsible for a stage of cosmic inflation in the early
Universe~\cite{Lerner:2009xg,Kahlhoefer:2015jma,Tenkanen:2016idg,Ema:2017ckf}
(see also~Ref.~\cite{Enqvist:2014zqa}).
A third possibility, which has been intensively studied in the literature,
is to associate the field $S$ with the dynamics of EW symmetry
breaking, in particular, in such a way that the EW phase transition
turns into a strong first-order phase transition~\cite{McDonald:1993ey,
Profumo:2007wc,Barger:2008jx,Espinosa:2011ax,Cline:2012hg,Alanne:2014bra}.
Similarly, the field $S$ might correspond to a pseudo-Nambu-Goldstone
boson (pNGB) in models that attribute the origin of the SM Higgs potential
to the spontaneous breaking of an approximate global symmetry.
In Ref.~\cite{Alanne:2017sip}, 
it was shown that, coupling such a pNGB singlet to RHNs, it is possible to connect
low-scale leptogenesis to EW symmetry breaking and Goldstone-Higgs models.
From a theoretical perspective, the singlet $S$ is motivated
by the fact that it allows one to ensure absolute stability of the EW
vacuum, as opposed to the metastability of the EW
vacuum in the pure standard model~\cite{Lebedev:2012zw,EliasMiro:2012ay}.
Finally, the extension of the SM by a real scalar singlet
represents an important experimental benchmark scenario for new-physics searches
at present and future colliders~\cite{Barger:2007im,Robens:2015gla,Martin-Lozano:2015dja,
Falkowski:2015iwa,Buttazzo:2015bka,No:2018fev,Buttazzo:2018qqp}. 


We stress that, despite this broad spectrum of possible applications,
the physical origin of the field $S$ is, in fact,
irrelevant for our purposes.
As we will see in the course of our analysis, leptogenesis will not
require us to specify an ultraviolet (UV) completion of our model.
Instead, it will suffice to regard the field $S$ as part
of a low-energy effective description
that descends from some unknown
dynamics at high energies.
In this sense, our analysis will follow a bottom-up
approach that promises to be compatible with a large
range of UV-complete models.


Our crucial observation is that the singlet $S$ is,
\textit{a priori}, always allowed to couple to
the heavy Majorana neutrinos in the seesaw sector,
$\mathcal{L} \supset SNN$.
This operator is renormalizable and trivially invariant under all
SM gauge symmetries. 
It explicitly breaks the accidental global lepton-number
symmetry of the SM, but the same is also true for
the large RHN Majorana masses in the seesaw Lagrangian.
We therefore argue that the singlet-neutrino Yukawa coupling $SNN$ is
likely to be present in any low-energy effective theory 
that contains a real scalar singlet and a set of sterile neutrinos.
This is a characteristic and fascinating consequence of combining two
popular SM extensions: the real-scalar-singlet extension
of the Higgs sector on the one hand and the type-I seesaw extension of the
neutrino sector on the other hand.
Note that, apart from Higgs portal couplings in the potential,
other renormalizable couplings between the singlet 
and SM degrees of freedom (DOFs) are forbidden by gauge invariance.


\begin{figure*}[t]
\centering
\begin{minipage}[c]{.27\textwidth}
  \begin{tikzpicture}[node distance=1cm]
    \coordinate[label=left:$N_2$] (v1);
    \coordinate[vertex, right = of v1] (v2);
    \coordinate[vertex, below right = of v2] (v3);
    \coordinate[vertex, above right = of v2] (v4);
    \coordinate[right = of v3, label=right:$\ell_\alpha$] (v5);
    \coordinate[right = of v4, label=right:$H$] (v6);
    \draw[mfermion] (v1) -- (v2);
    \draw[rscalar] (v4) -- node[label=above left:$S$] {} (v2);
    \draw[mfermion] (v3) -- node[label=below left:$N_1$] {} (v2);
    \draw[scalar] (v3) -- node[label=right:$H$] {} (v4);
    \draw[scalar] (v4) -- (v6);
    \draw[fermion] (v3) -- (v5);
  \end{tikzpicture}
\end{minipage}
  \hspace{15mm}
\begin{minipage}[c]{.33\textwidth}
  \begin{tikzpicture}[node distance=1cm]
    \coordinate[label=left:$N_2$] (v1);
    \coordinate[vertex, right = of v1] (v2);
    \coordinate[vertex, right = of v2] (v3);
    \coordinate[vertex, right = of v3] (v4);
    \coordinate[below right = of v4, label=below right:$H$] (v5);
    \coordinate[above right = of v4, label=above right:$\ell_\alpha$] (v6);
    \draw[mfermion] (v1) -- (v2);
    \draw[mfermion] (v2) arc(180:90:.56) node[label=above:$N_1$]{} arc(90:0:.56);
    \draw[rscalar] (v2) arc(-180:-90:.56) node[label=below:$S$]{} arc(-90:0:.56);
    \draw[mfermion] (v3) -- node[label=above:$N_2$] {} (v4);
    \draw[scalar] (v4) -- (v5);
    \draw[majfermion] (v4) -- (v6);
  \end{tikzpicture}
\end{minipage}
\caption{\label{fig:CPasymmetryMod}Scalar-singlet-mediated one-loop diagrams
that enhance the amount of $CP$ violation in the decay $N_2 \to \ell_\alpha H$.}
\end{figure*}
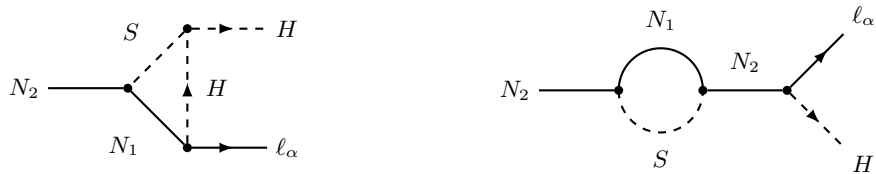


The operator $SNN$ has important consequences for leptogenesis. 
In this paper, we will study its impact assuming
a \textit{minimal} singlet sector that consists of only two
heavy neutrinos $N_{1,2}$ and one real scalar $S$.
This restriction will help us simplify our analysis, but at the same time
still manage to capture all relevant effects.
At tree level, the presence of a light scalar $S$ then leads to a
new $N_2$ decay channel, $N_2 \rightarrow N_1 S$,
which causes the $N_2$ number density to depart more strongly
from its would-be value in thermal equilibrium.
At the same time, the new $N_2$ decay channel also increases the amount
of $CP$ violation in $N_2$ decays at loop level.
Together, these two effects enhance the efficiency of leptogenesis,
such that the BAU can be successfully generated in $N_2$ decays
for RHN masses even below the TeV scale.
The general principle behind this mechanism has been described for
the first time by Le~Dall and Ritz~\cite{Dall:2014nma}.%
\footnote{The first discussion of $N_2$-dominated leptogenesis in the
standard type-I seesaw model can be found in Ref.~\cite{DiBari:2005st}.}
In this paper, we will build upon this earlier work and extend it
in several directions.
An important novel aspect of our analysis is that we perform
a systematic scan of parameter space, which allows us
to determine the dependence of the final asymmetry
on the choice of parameter values in the neutrino sector.
In particular, we derive a semianalytical fit function
that allows us to reproduce the exact numerical result
with high precision [see Eqs.~\eqref{eq:finalBasymmetry} and \eqref{eq:fit}].
Moreover, we make sure that we only consider parameter values
in the neutrino sector that are consistent with
the most recent low-energy data on neutrino oscillations.
This is accomplished by employing the Casas-Ibarra (CI) parametrization
for the RHN Yukawa couplings to SM lepton-Higgs pairs~\cite{Casas:2001sr}.
 

The rest of this paper is organized as follows.
In Sec.~\ref{sec:framework}, we will outline the characteristics
of our model and discuss how the scalar $S$ affects
some of the key quantities in the description of leptogenesis.
In Sec.~\ref{sec:analytics}, we will then turn to the modified Boltzmann
equations and present our semianalytical fit function for the BAU.
In Sec.~\ref{sec:ScalarConst}, we will discuss further theoretical
and phenomenological constraints on our scenario and highlight 
the viable regions in parameter space.
Sec.~\ref{sec:conclusions} contains our conclusions, and 
in Appendix~\ref{app:formulas}, we collect 
some lengthy formulae.


\section{Real-scalar-singlet extension of the type-I seesaw mechanism}
\label{sec:framework}


\subsection{Couplings and masses in the Lagrangian}


The starting point of our analysis is the type-I
seesaw model featuring two heavy neutrinos $N_{1,2}$ and supplemented
by the interactions of the real scalar singlet, $\widetilde{S}$,
\begin{align}
\label{eq:LagHE}
- \mathcal{L} & \supset
\left[\widetilde{h}_{\alpha i}\,\ell_\alpha  \widetilde{N}_i H
+ \frac{1}{2} \left(\widetilde{M}_{ij} + \widetilde{\alpha}_{ij} \widetilde{S}\right)
\widetilde{N}_i \widetilde{N}_j
+ \textrm{H.c.}\right]
\nonumber \\
& + V\big(H,\widetilde{S}\,\big) \,; \qquad i,j = 1,2 \,; \quad \alpha = e,\mu,\tau \,.
\end{align}
Here, we use a tilde above every quantity whose definition
depends on the configuration of vacuum expectation values (VEVs)
in the scalar sector.
In Eq.~\eqref{eq:LagHE}, all the quantities equipped with a tilde
are defined at high energies, where the scalar VEVs in our model are
assumed to vanish.
These fields and parameters should thus be considered as the fundamental
input quantities of our model.
The fields $H = \left(H^+,H^0\right)^T$ and
$\ell_\alpha = \left(\nu_\alpha,\alpha_L\right)^T$ correspond to the SM Higgs doublet
and the three SM left-handed charged-lepton doublets, respectively.
We denote the usual RHN-lepton-Higgs Yukawa matrix by $\widetilde{h}$,
and $\widetilde{M}$ is a matrix of RHN input masses
whose dynamical origin is left unspecified for the purposes of this work.
The matrix $\widetilde{\alpha}$ characterizes the strength of 
the novel singlet-RHN Yukawa interactions.
The scalar $\widetilde{S}$ will only be able to modify the dynamics
of leptogenesis if the two matrices $\widetilde{M}$ and $\widetilde{\alpha}$
are not proportional to each other.
That is, only if $\widetilde{M}$ and $\widetilde{\alpha}$ cannot be diagonalized
simultaneously, will the scalar $\widetilde{S}$ induce flavor-changing neutral-current
interactions among the heavy neutrinos that affect the efficiency of leptogenesis.
We stress once more that both the RHN mass terms as well as the singlet-neutrino
Yukawa interactions explicitly violate global lepton number.


The scalar potential, $V$, in Eq.~\eqref{eq:LagHE} encompasses
the SM Higgs potential as well as a number of terms involving
the new scalar field, $\widetilde{S}$.
We emphasize that the requirement of successful leptogenesis only leads
to weak constraints on the overall shape of the scalar potential. 
The most important requirement is that the singlet scalar must share a
trilinear coupling with two SM Higgs fields.  
Together with the new Yukawa interactions parametrized by $\widetilde{\alpha}$,
this operator leads to a new source of $CP$ violation in $N_2$ decays at
the one-loop level (see Fig.~\ref{fig:CPasymmetryMod}).
One possibility is that the trilinear coupling is simply generated in the
UV completion together with all other couplings of the scalar singlet $S$.
Alternatively, one may assume that the singlet is odd under a $\mathbb{Z}_2$ symmetry of the 
scalar potential, and the trilinear coupling is generated via the Higgs-portal operator, 
$\widetilde{S}^2\left|H\right|^2$, as a consequence of spontaneous symmetry breaking.
In the rest of this paper, we will pursue this second, $\mathbb{Z}_2$-symmetric
option for illustrative purposes.
At the same time, we emphasize that the results for leptogenesis
do not depend on this choice, but are valid for more generic scalar
potentials containing a trilinear $\widetilde{S}\left|H\right|^2$ coupling.
The restriction to the $\mathbb{Z}_2$-symmetric case, however, has several advantages.
First of all, it comes with a smaller
number of free input parameters in the renormalizable part of the Lagrangian.
This will facilitate our discussion of the experimental and theoretical
bounds on our model in Sec.~\ref{sec:ScalarConst}.
But in addition, a discrete $\mathbb{Z}_2$ symmetry may also provide
us with some intuition regarding the smallness of the Yukawa 
couplings, $\widetilde{\alpha}$, which break this global $\mathbb{Z}_2$ symmetry explicitly.
The $\mathbb{Z}_2$ symmetry in the scalar potential can therefore be
at most an \textit{approximate} symmetry that is explicitly broken by the
singlet-neutrino Yukawa interactions.
Conversely, this means that small values of the $\widetilde{\alpha}$ Yukawa couplings
are symmetry-protected and hence natural in the sense of 't Hooft~\cite{tHooft:1979rat}.


Let us now suppose that the singlet field in Eq.~\eqref{eq:LagHE}
does obtain a nonzero VEV, $v_S$, at some high temperature
above the EW phase transition, $T_S \gg T_{\rm ew}$,
and that leptogenesis occurs at energies in between these temperatures.  
It is then convenient to shift the scalar singlet
by its VEV, $\widetilde{S} \rightarrow v_S + S$, and to perform a
transformation on the heavy neutrinos in order to
diagonalize their mass matrix $\widetilde{M}$.
After these two steps, the Lagrangian reads
\begin{align}
\label{eq:LagLep}
- \mathcal{L} & \supset
\left[h_{\alpha i}\,\ell_\alpha  N_i H
+ \frac{1}{2} \left(\delta_{ij}\,M_j + \alpha_{ij}\,S\right) N_i N_j
+ \textrm{H.c.}\right]
\nonumber \\
& + V\left(H,S\right) \,,
\end{align}
where we dropped all tildes to indicate that this Lagrangian is now
valid at energies below the $\mathbb{Z}_2$-symmetry breaking scale.
This Lagrangian encodes the RHN interactions that are relevant
for leptogenesis.
It is interesting to note that the masses and couplings
in Eq.~\eqref{eq:LagLep} still do not fully coincide with
the respective parameters at low energies.
The reason for this is that the singlet VEV, $v_S$, is expected to receive corrections
during the EW phase transition, $v_S \rightarrow v_S^0$, which
necessitates yet another shift in the scalar singlet, $S \rightarrow v_S^0-v_S + s$,
as well as yet another diagonalization of the RHN mass matrix.%
\footnote{These corrections are related to $\alpha_{ij}$ and therefore
small in our scenario, so we will neglect them in the analysis of leptogenesis.}

After EW symmetry breaking, the final low-energy Lagrangian in unitary
gauge obtains the following form,
\begin{align}
\label{eq:LagLE}
- \mathcal{L} & \supset
\left[\frac{h_{\alpha i}^0}{\sqrt{2}}\,\nu_\alpha  N_i^0 h
+ \frac{1}{2} \left(\delta_{ij}\,M_j^0 + \alpha_{ij}^0\,s\right) N_i^0 N_j^0
+ \textrm{H.c.}\right]
\nonumber \\
& + V\left(h,s\right) \,,
\end{align}
where $h$ denotes the physical SM Higgs boson with a mass of $125\GeV$.
The Lagrangian in Eq.~\eqref{eq:LagLE} sets the stage for the type-I seesaw mechanism.
Upon integrating out the heavy sterile neutrinos, the light SM
neutrinos acquire a Majorana mass matrix, $m_\nu$, of the following
form~\cite{Minkowski:1977sc,Yanagida:1979as,Yanagida:1980xy,GellMann:1980vs,Mohapatra:1979ia},
\begin{align}
m_\nu = - m_D\,D_N^{-1}\,m_D^T \,,
\end{align}
where $m_D = h^0/\sqrt{2}\,v_{\rm ew}$ denotes the neutrino Dirac
mass matrix and $D_N = \textrm{diag}\left(M_1^0,M_2^0\right)$
is the diagonal heavy-neutrino Majorana mass matrix. 
The light-neutrino Majorana mass matrix, $m_\nu$, can
be diagonalized by acting on it with a unitary transformation,
$D_\nu = U^T\,m_\nu\,U$.
In the flavor basis where the charged-lepton Dirac
mass matrix is diagonal, the matrix $U$ coincides
with the Pontecorvo-Maki-Nakagawa-Sakata (PMNS) lepton mixing
matrix~\cite{Pontecorvo:1957qd,Maki:1962mu}.
According to the CI parametrization~\cite{Casas:2001sr},
the Yukawa matrix, $h^0$, can then be partially reconstructed based on
the information contained in $U$, $D_\nu$, and $D_N$:
\begin{align}
\label{eq:CIP}
\frac{h^0}{\sqrt{2}} = \frac{i}{v_{\rm ew}}\,U^*D_\nu^{1/2}R\,D_N^{1/2} \,.
\end{align}
For only two heavy sterile neutrinos, $R$ is a
complex $3\times2$ rotation matrix that satisfies $R^TR = \mathbb{1}$
and that can be parametrized in terms of one complex rotation angle $z$.
Besides that, the right-hand side of Eq.~\eqref{eq:CIP} also depends
on the mixing angles $\theta_{12}$, $\theta_{13}$, and $\theta_{23}$, 
the phases $\delta_{\rm CP}$ and $\alpha_{\rm CP}$
in the PMNS matrix $U$, the mass splittings
$\Delta m_{\rm sol}^2$ and $\Delta m_{\rm atm}^2$ in the light-neutrino mass
spectrum, and the mass eigenvalues $M_1^0$ and $M_2^0$ of the
two heavy sterile neutrinos.
In this paper, we shall neglect all flavor effects in the computation
of the final BAU for simplicity.
As we will see, the dynamics of leptogenesis will thus be controlled by the entries
of the Hermitian matrix $h^{0\dagger} h^0$,
\begin{align}
\label{eq:hdagh}
\frac{h^{0\dagger} h^0}{2} = \frac{1}{v_{\rm ew}^2}\,
D_N^{1/2}R^\dagger D_\nu R\,D_N^{1/2} \,,
\end{align}
which is independent of the PMNS matrix.
In our numerical analysis in Sec.~\ref{sec:analytics}, we will use the
best-fit values for $\Delta m_{\rm sol}^2$, and $\Delta m_{\rm atm}^2$ according to the most
recent global-fit analysis presented by the NuFIT collaboration~\cite{Esteban:2018azc}.
The RHN mass eigenvalues $M_1^0$ and $M_2^0$ as well as the
complex rotation angle $z$ will be treated as free input parameters.


\subsection{\texorpdfstring{\boldmath{$CP$}}{CP} asymmetry in RHN decays}


The decay of the heavy Majorana neutrinos into SM lepton-Higgs
pairs, $N_i \rightarrow \ell_\alpha H,\,\ell_\alpha^\dagger H^\dagger$,
violates $CP$ invariance.
The amount of $CP$ violation in these decays is conveniently
measured by the $CP$-asymmetry parameters
\begin{align}
\varepsilon_i = \frac{\sum_\alpha\big[
  \Gamma(N_i \rightarrow\ell_\alpha H)
- \Gamma(N_i \rightarrow\ell_\alpha^\dagger H^\dagger)\big]}
{\sum_\alpha\big[
  \Gamma(N_i \rightarrow\ell_\alpha H)
+ \Gamma(N_i \rightarrow\ell_\alpha^\dagger H^\dagger)\big]} \,.
\end{align}
In the standard type-I seesaw model without any extra scalar DOFs,
evaluating this expression results in~\cite{Covi:1996wh}
\begin{align}
\varepsilon_i^0 = \frac{1}{8\pi\left(h^\dagger h\right)_{ii}} \sum_{j \neq i}\,
\textrm{Im}\left[\left(h^\dagger h\right)_{ji}^2\right] \,
\mathcal{F}\left(\frac{M_j}{M_i}\right) \,, 
\label{eq:CPasymmetry}
\end{align}
with the loop function $\mathcal{F}$ being defined as follows:
\begin{align}
\mathcal{F}\left(x\right) = x \left[
1 + \left(1 + x^2\right)\ln\left(\frac{x^2}{x^2 + 1}\right) - \frac{1}{x^2-1} \right] \,.
\end{align}
This result needs to be compared to the $CP$-asymmetry parameters
$\varepsilon_{1,2}$ in the presence of the scalar singlet, $S$.
The decay $N_1 \rightarrow N_2\,S$ is kinematically forbidden, which
is why the parameter $\varepsilon_1^0$ remains unchanged.
The parameter $\varepsilon_2^0$, however, receives further contributions
from the one-loop vertex (v) and self-energy (s) diagrams
in Fig.~\ref{fig:CPasymmetryMod},
\begin{align}
\label{eq:eps12}
\varepsilon_1 = \varepsilon_1^0 \,, \quad
\varepsilon_2 = \varepsilon_2^0 + \varepsilon_2^{\rm v} + \varepsilon_2^{\rm s} \,.
\end{align}
An explicit computation of the extra terms yields~\cite{Dall:2014nma}
\begin{subequations}
\label{eq:CPextra}
\begin{align}
8\pi\left(h^\dagger h\right)_{22}\,\varepsilon_2^{\rm{v}} = &\,
\Im{\left(h^\dagger h\right)_{12}\beta_2\,\alpha_{21}}
\mathcal{F}^{\rm{v}}_{21,\,R} \nonumber\\
+ &\,\Im{\left(h^\dagger h\right)_{12}\beta_2\,\alpha_{21}^*}
\mathcal{F}^{\rm{v}}_{21,\, L} \,, \\
8\pi\left(h^\dagger h\right)_{22}\,\varepsilon_2^{\rm{s}} = &\,
\Im{\left(h^\dagger h\right)_{12}\alpha_{21}\,\alpha_{11}}
\mathcal{F}^{\rm{s}}_{211,\,RR} \nonumber\\
+ &\,\Im{\left(h^\dagger h\right)_{12}\alpha_{21}^*\,\alpha_{11}}
\mathcal{F}^{\rm{s}}_{211,\,RL} \nonumber\\
+ &\,\Im{\left(h^\dagger h\right)_{12}\alpha_{21}\,\alpha_{11}^*}
\mathcal{F}^{\rm{s}}_{211,\,LR} \nonumber\\
+ &\,\Im{\left(h^\dagger h\right)_{12}\alpha_{21}^*\,\alpha_{11}^*}
\mathcal{F}^{\rm{s}}_{211,\,LL} \,,
\end{align}
\end{subequations}
where $\beta_2 = \mu / M_2$ is related to the dimensionful
coupling $\mu$ of the trilinear $S\left|H\right|^2$
operator in the scalar potential,
\begin{align}
V\left(H,S\right) \supset \mu\,S\left|H\right|^2 \,.
\label{eq:SHH}
\end{align}
General expressions for the  
loop functions in Eq.~\eqref{eq:CPextra}
can be found in Appendix~\ref{app:formulas}.
The indices $R$ and $L$ indicate
whether a given contribution to $\varepsilon_2$ stems
from the interaction of the scalar field, $S$, with a
pair of right-chiral or left-chiral neutrino spinor fields, respectively. 


\begin{figure}[t]
\centering
\includegraphics[width=0.5\textwidth]{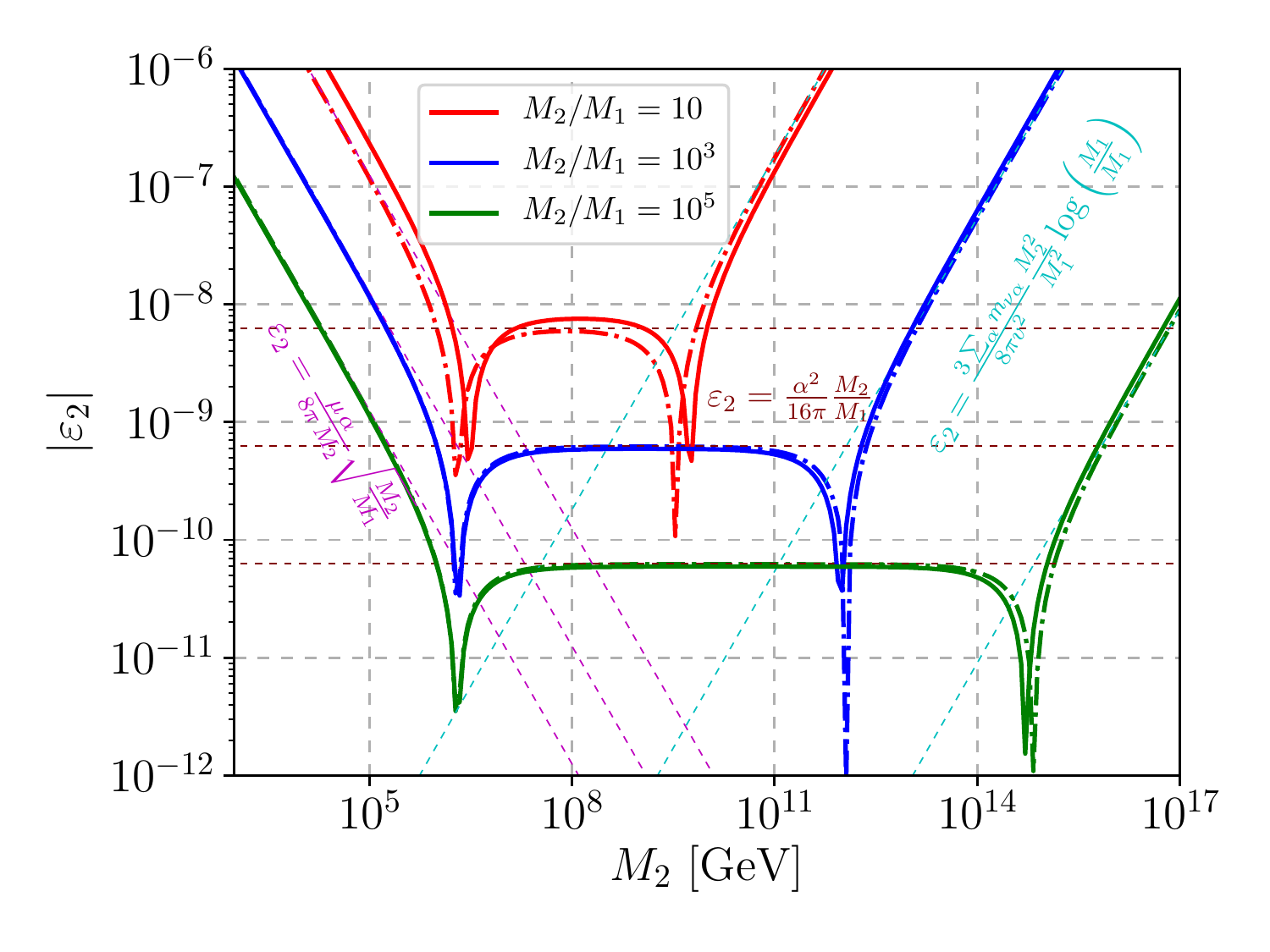}
\caption{\label{fig:CPasymmetry_estimate}
$CP$-asymmetry parameter $\varepsilon_2$ in Eq.~\eqref{eq:CPasymmetry}
as a function of the heavy RHN mass.
Here, ${\alpha_{ij} = 10^{-3}}$, ${\mu = 1\TeV}$, and ${m_S \ll M_2}$.
The solid lines show the exact result; the dashed lines indicate the
estimates in Eq.~\eqref{eq:CPextra2} (see also Ref.~\cite{Dall:2014nma}); and
the dot-dashed lines represent the sums of the respective estimates.
The exact asymmetry changes sign depending on the signs of the couplings and CI angle.}
\end{figure}


The expressions in Eq.~\eqref{eq:CPextra} illustrate
the importance of the off-diagonal singlet-neutrino
Yukawa coupling, $\alpha_{21}$.
Only if this coupling is nonzero 
will the scalar field $S$ mediate flavor-changing neutral-current interactions
among the heavy neutrinos and hence lead to an
enhanced $CP$ asymmetry in $N_2$ decays.
As we will see in Sec.~\ref{sec:analytics}, an important feature
of our scenario is that this gain in $CP$ asymmetry
is not counteracted by a correspondingly stronger washout.
The reason for this is that the decay $N_2 \rightarrow N_1 S$ does
not involve the usual SM lepton-Higgs pairs and thus does not
increase the strength of the usual asymmetry washout by inverse decays.
This needs to be contrasted with the situation in standard
thermal leptogenesis, where both the amount of $CP$ asymmetry and
the washout efficiency are controlled by the same combination of
couplings, i.e. by the same products of Yukawa couplings
of the form $h_{\alpha i}^* h_{\beta j}^{\vphantom{*}}$.


For a sufficiently hierarchical RHN mass spectrum,
the $CP$-asymmetry parameters $\varepsilon_2^{\rm{v}}$ and $\varepsilon_2^{\rm{s}}$
in Eq.~\eqref{eq:CPextra} can be considerably simplified.
For $M_2 \gg M_1$, one obtains
\begin{align}
\begin{split}
\varepsilon_2^{\rm{v}} & \approx - \frac{\left|\beta_2\,\alpha_{21}\right|}{8\pi}
\sqrt{\frac{M_1}{M_2}}\left(1-\frac{m_{S}^2}{M_2^2}\right) \,, \\
\varepsilon_2^{\rm{s}} & \approx \frac{\left|\alpha_{21}\,\alpha_{11}\right|}{16\pi}
\sqrt{\frac{M_1}{M_2}}\left(1 - \frac{m_{S}^2}{M_2^2}\right)^2.
\end{split}
\label{eq:CPextra2}
\end{align}
In Fig.~\ref{fig:CPasymmetry_estimate}, we compare these estimates (dashed lines)
with the corresponding exact expressions in Eqs.~\eqref{eq:CPextra} (solid lines).
As evident from this figure, the new contributions to the $CP$
asymmetry are dominant for comparatively low $N_2$ masses. 
The new vertex contribution $\varepsilon_2^{\rm{v}}$, e.g., leads
to a strong enhancement of the total $CP$ asymmetry
for $N_2$ masses less than $\mathcal{O}\left(10^6\right)\GeV$.
Unfortunately, this increase in the $CP$ asymmetry will not enable
us to lower the energy scale of leptogenesis to arbitrarily small
RHN masses.
As we will discuss in Sec.~\ref{sec:Scattering}, 2-to-2 scattering processes
will 
eventually give rise to new sources of washout and thus
limit the efficiency of leptogenesis. 
From Eq.~\eqref{eq:CPextra2}, one can also see that at some point a larger RHN mass scale will lower $\varepsilon_2^{\rm{v}}$ below $\varepsilon_2^{\rm{s}}$, and the contribution becomes independent of $\mu$. Finally, the $CP$ asymmetry will revert to grow proportional to the RHN masses, and we enter the standard type-I regime.


\section{Boltzmann equations and a semianalytical solution}
\label{sec:analytics}


To gain a physical understanding of the scalar-assisted leptogenesis induced by $N_2$ decays, we consider a coupled set of classical Boltzmann equations for the normalized number densities, $N_{N_i}$. These capture the decays and inverse decays of RHNs into leptons and the Higgs doublet, as well as the corresponding processes for the decay ${N_2 \to N_1\, S}$. Furthermore, we include the dominant washout 
effect induced by the $\Delta L = 2$ scatterings $N_iN_j\to HH$ mediated by the singlet. The relevant equations read~\cite{Dall:2014nma}
\begin{subequations}\label{eq:BoltzmannEqs}\allowdisplaybreaks
\begin{align}
    \begin{split}
	\frac{\mathrm{d} N_{N_2}}{\mathrm{d} z} =& - (D_2 + D_{21})\, \Delta_{N_2} + D_{21}\, \Delta_{N_1}\\
	    &-\Delta_{N_1N_2}S_{N_1N_2\rightarrow HH}\\
	    &-\Delta_{N_2N_2}S_{N_2N_2\rightarrow HH}
    \end{split} \label{eq:Boltzmann2}
    \\
    \begin{split}
	\frac{\mathrm{d} N_{N_1}}{\mathrm{d} z} =& - (D_1 + D_{21})\, \Delta_{N_1} + D_{21}\, \Delta_{N_2} \\
	    &-\Delta_{N_1N_2}S_{N_1N_2\rightarrow HH}\\
	    &-\Delta_{N_1N_1}S_{N_1N_1\rightarrow HH},
    \end{split}\label{eq:Boltzmann1}\\
  \frac{\mathrm{d} N_{B-L}}{\mathrm{d} z} =& - \sum_{i=1}^2 \varepsilon_i\, D_i \Delta_i - W N_{B-L},\label{eq:BoltzmannBL}
\end{align}
\end{subequations}
with the temperature-dependent quantities
\begin{subequations}\label{eq:Boltz_defs}\allowdisplaybreaks
\begin{gather}
    \Delta_{N_i}(z) \equiv \frac{N_{N_i}(z)}{N_{N_i}^{\text{eq}}(z)} - 1\,,\\
    \Delta_{N_iN_j} \equiv \frac{N_{N_i}N_{N_j}}{N_{N_i}^{\mathrm{eq}}N_{N_j}^{\mathrm{eq}}}-1 \,,\\
  N_{N_i}^\text{eq}(z) = \frac{z_i^2}{2}\,\mathcal{K}_2(z_i)\\
  D_i (z)= K_i\,z \,\frac{\mathcal{K}_1(z_i)}{\mathcal{K}_2(z_i)} N_{N_i}^{\text{eq}}(z)\, ,\\
  D_{21}(z) = K_{21} \,z\,\frac{\mathcal{K}_1(z_2)}{\mathcal{K}_2(z_2)} N_{N_2}^{\text{eq}}(z)\,, \\
  W(z) = \sum_i \frac{1}{4}\,K_i\,z_i^3\,\mathcal{K}_1(z_i) .
\end{gather}
\end{subequations}
Here, $\mathcal{K}_a$ denotes the $a$-th modified Bessel function of the second kind, and $z_i \equiv M_i / T$ is the inverse dimensionless temperature for the $i$-th RHN, with $z \equiv z_1$. The relevant decay parameters read, respectively,
\begin{equation}\label{eq:decay_parameter}
K_i \equiv \frac{\Gamma(N_i\to L H)}{H(T=M_i)}
= \sqrt{\frac{45}{64 \pi^5 g_*}} \frac{M_{\rm Pl}}{v_{\rm ew}^2}
\left( \mathcal{M}_\nu^R \right)_{ii}\,,
\end{equation}
with the Hubble rate $H = \sqrt{8 \pi^3 g_*/90}\:T^2/M_\text{Pl}$. The quantity $\mathcal{M}_\nu^R$ emerges from the CI parametrization for the neutrino Yukawa matrix, Eq.~\eqref{eq:hdagh}, and is given by
\begin{equation}
	\mathcal{M}_\nu^R \equiv R^\dagger D_{\nu} R,
\end{equation}
where $R$ is the aforementioned arbitrary complex matrix that satisfies $R^T R =\mathbb{1}$. Note that the decay parameters $K_i$ are independent of the RHN masses and that, for two RHNs, their values are bounded to be larger than $K_i \gtrsim 8$, i.e.,~we are always in the regime of strong washout. The decay parameter of the new decay channel reads
\begin{equation}
	K_{21}  \equiv \frac{\Gamma(N_2\to N_1 S)}{H(T=M_2)},
\end{equation}
where the decay width for $N_2 \to N_1 S$ is given by~\cite{Dall:2014nma}
\begin{equation}
	\Gamma(N_2 \to N_1 S) = \frac{\left|\alpha_{12}\right|^2 M_2}{16 \pi} \left[ \left(1+ r_{12}\right)^2 - \sigma_2 \right] \sqrt{\delta_{12}},
\end{equation}
with 
\begin{equation}
    \begin{split}
    &r_{ij} \equiv (M_i / M_j)^2,\quad  \sigma_i \equiv m_S^2 / M_i^2\,,\\ &\delta_{ij} \equiv (1-r_{ij} -\sigma_j)^2 - 4\,r_{ij} \sigma_j.
    \end{split}
\end{equation}
Typically, the new decay parameter is very large, e.g.,~$K_{21} \sim 10^6$ for $M_2 = 1\TeV$ and $\alpha_{12} = 10^{-3}$. This is crucial for the RHN number densities to display a significant deviation from their equilibrium form.
We will discuss the impact of the scattering terms, $S_{N_iN_j\to HH}$, and
give their explicit expressions in Sec.~\ref{sec:Scattering}. 
For the rest of this section, we will assume that all scattering
terms in Eqs.~\eqref{eq:BoltzmannEqs} are negligible.


\begin{figure}[t]
\centering
    \includegraphics[width=.5\textwidth]{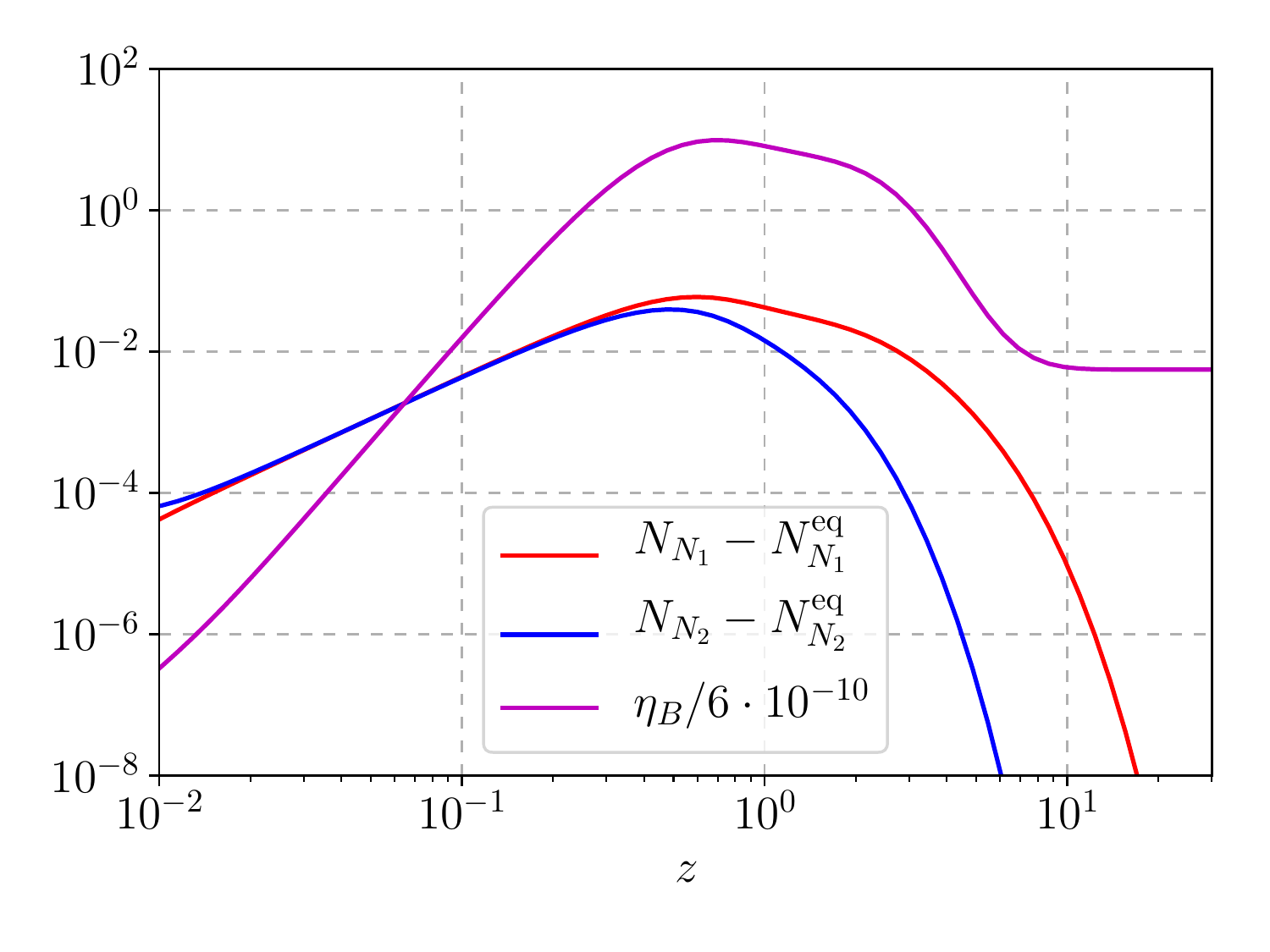}
\caption{\label{fig:Example}Sample evolution of RHN number density and baryon asymmetry as a function of $z \equiv z_1 = M_1 / T$. For this scenario, we fixed the parameters as follows: $M_1 = 2 \TeV$, $M_2 = 6 \TeV$, $m_S = \mu = 500\GeV$, $m_H = 0$, and $\alpha_{ij} = 10^{-3}$. The CI rotation angle $z$ is fixed such that $K_1 = 12$, $K_2 = 51$.}
\end{figure}


Solving this set of coupled differential equations, we see that the genesis of a lepton number asymmetry proceeds in two steps; cf.~Fig.~\ref{fig:Example}: First, the $N_2$ decays into leptons and anti-leptons build up a large lepton or $B-L$ asymmetry around  temperatures $z_2 \lesssim 1$. This is driven mostly by the enhanced $CP$ asymmetry due to the new decay channels. Subsequently, once $N_1$ drops out of thermal equilibrium, around $z_1 \sim 1$, the analogous decay for $N_1$ occurs; however, due to the small $CP$ asymmetry, this effect is negligible for $N_1$ masses $\lesssim 10^6\GeV$~\cite{Dall:2014nma}. At the same time, inverse $N_1$ decays wash out the asymmetry previously built up by the decay of $N_2$, reducing the final $B-L$ asymmetry exponentially, as one can infer from the last term in Eq.~\eqref{eq:BoltzmannBL} and the rapid decrease of the purple curve in Fig.~\ref{fig:Example} before becoming constant. The overall magnitude of the washout is, however, smaller than in the standard type-I seesaw framework, as a deviation of the $N_1$ number density from thermal equilibrium immediately leads to a comparable deviation of the $N_2$ number density from its equilibrium value by virtue of the large $K_{21}$. Thus, strong washout is always accompanied by more $CP$ violating decays, enhancing the lepton asymmetry and counteracting the washout.


\begin{figure*}[t]
	\begin{center}
	\includegraphics[width=.9\textwidth]{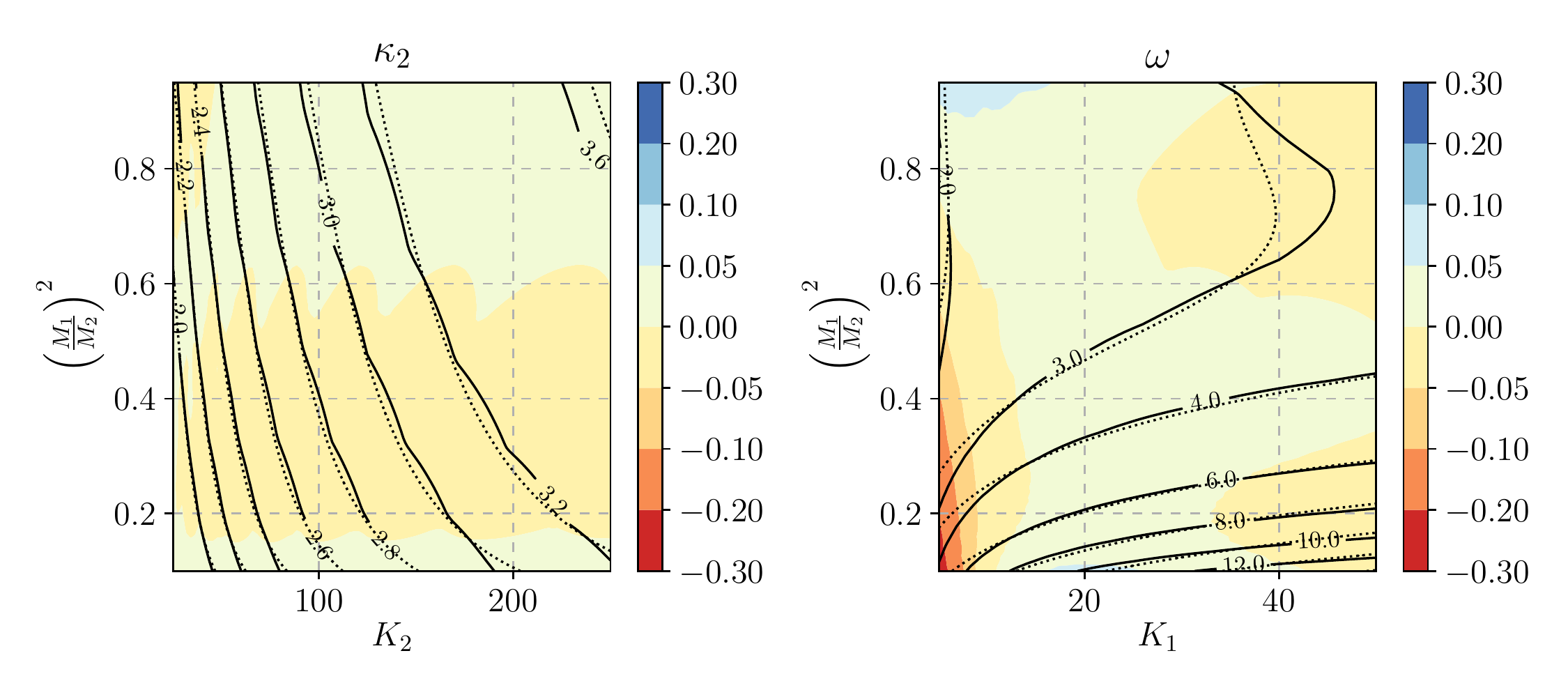}
	\end{center}
    \caption{\label{fig:fit}Comparison of the efficiency factor $\kappa_2$ and the washout exponent $\omega$ for the numerical scan (solid lines) and the fit functions presented in Eqs.~(\ref{eq:finalBasymmetry}, \ref{eq:fit}) (dashed lines). The relative deviation is shown in color coding and is less than 10\% for most of the parameter values of interest.}
\end{figure*}


The final $B-L$ asymmetry is subsequently converted into the baryon asymmetry via EW sphalerons, which become inefficient below a temperature $T_\text{sph} = 131\GeV$~\cite{DOnofrio:2014rug}, and therefore the relevant asymmetry needs to be generated at temperatures above $T_\text{sph}$. Note, however, that an asymmetry generated \emph{just before} sphaleron interactions freeze-out will persist without being subject to washout (which affects only the lepton sector). This has already been used in Ref.~\cite{Dall:2014nma} to find low-mass scenarios with $M_1 < T_\mathrm{sph} \lesssim M_2$. The final baryon-to-photon ratio is given by ${\eta_B \simeq 0.01 \, N_{B-L}(z=\infty)}$, which accounts for the sphaleron conversion and entropy production after the generation of the lepton asymmetry and is already included in Fig.~\ref{fig:Example}. However, we remark that, while the example on display highlights the distinct regimes, the resulting asymmetry is two orders of magnitude lower than the measured value. This indicates that some additional source of $CP$ violation is needed\,---\,limited by scattering processes; cf.~Sec.~\ref{sec:Scattering}. Alternatively, the masses must be of similar magnitude in order to suppress the washout. 


By numerical integration of Eqs.~\eqref{eq:BoltzmannEqs}, one readily obtains a value for $N_{B-L}(z=\infty)$; however, the physical insight is limited due to the extended parameter space of the model. The subsequent subsections are dedicated to the development of an analytical understanding of these mechanisms and comparing them to the exact, numerical results including also the scattering terms. Eventually, we will use these results to identify parameter regions that successfully explain the observed baryon asymmetry.


\subsection{A semianalytical fit}


Let us try to simplify the treatment of leptogenesis by finding an approximation to the exact solution of the Boltzmann equations in Eq.~\eqref{eq:BoltzmannEqs}. As we observed in the scenario shown in Fig.~\ref{fig:Example}, there was not enough $CP$ violation and/or too much washout. While the $CP$ asymmetry can always be increased by raising $\mu$, cf.~Eqs.~\eqref{eq:CPextra} and~\eqref{eq:CPextra2}, eventually $\Delta L = 2$ scattering processes will become important and reduce the final asymmetry by increasing the washout. Thus, we seek to reduce the washout by reducing the mass splitting of the RHNs. As long as the $\Delta L = 2$ scattering processes are irrelevant, the final baryon asymmetry will be given by~\cite{Buchmuller:2004nz}
\begin{equation}\label{eq:finalBasymmetry}
	\eta_B \simeq 0.01 \, \kappa_2(K_2, r) \, \varepsilon_2 \times e^{- \omega(K_1, r)}\,,
\end{equation}
where $r \equiv r_{12} = (M_1 / M_2)^2$. This holds true, even in the non-hierarchical regime, because only the $N_2$ decays display significant $CP$ violation via Eqs.~\eqref{eq:CPextra}, and at the same time only $N_1$ decays wash out the asymmetry. While simple approximations assuming a large hierarchy between the RHN masses yield $\omega = 3\pi/8\,K_1$ and $\kappa_2 = 2 / z_B(K_2)$ with $z_B(K_2) = 2 + 4\,{K_2}^{0.13}\, e^{-2.5/K_2}$~\cite{Buchmuller:2004nz}, these will become more complicated as the masses are of the same order\,---\,however, non-degenerate (see also Ref.~\cite{Blanchet:2006dq}, which studied a similar setting for the type-I seesaw case). Allowing these functions to take a more general form, we find for the efficiency factor, $\kappa_2$, and the washout function, $\omega$, a good fit to be given by:
\begin{subequations}\label{eq:fit}
\begin{align}
      \kappa_2(K_2, r) =&\, \frac{2}{z_B(K_2)}\,  
      \frac{K_2^{0.44} \, r^{-0.2}}{0.3\,K_2 + 1.6\,r^{2.6}}\,,\\
      \omega(K_1, r) =&\, \frac{3 \pi}{8} \Big[- K_1^{0.5}(1.4\,r^2 - 2.1\,r + 0.8)\,\ln(r) \nonumber\\
      &\hspace{7mm}+1.2\, K_1^{0.2} \Big],
\end{align}
\end{subequations}
which are functions of the decay constants, $K_{1,\, 2}$, and the mass ratio $r$. In Fig.~\ref{fig:fit}, we show in solid contour lines the numerically calculated efficiency factor and washout exponent, which are determined in terms of the maximal $B-L$ asymmetry, $N^\mathrm{max}_{B-L} = \max_{z} N_{B-L} (z)$, as $\kappa_2 = N^\mathrm{max}_{B-L} / \varepsilon_2 \ge 0$ and $\omega = \ln N^\mathrm{max}_{B-L}/N_{B-L}(\infty) \ge 0$. These are compared to the above fit functions shown in dashed lines. The colors indicate the relative deviation of the numerically determined efficiency and washout compared to the fits, and we find that they provide a good approximation for the values of $K_{1,\, 2}$ and $r$ of interest to us with relative deviations not larger than about 10\%.


\begin{figure*}[t]
\centering
\includegraphics[width=\textwidth]{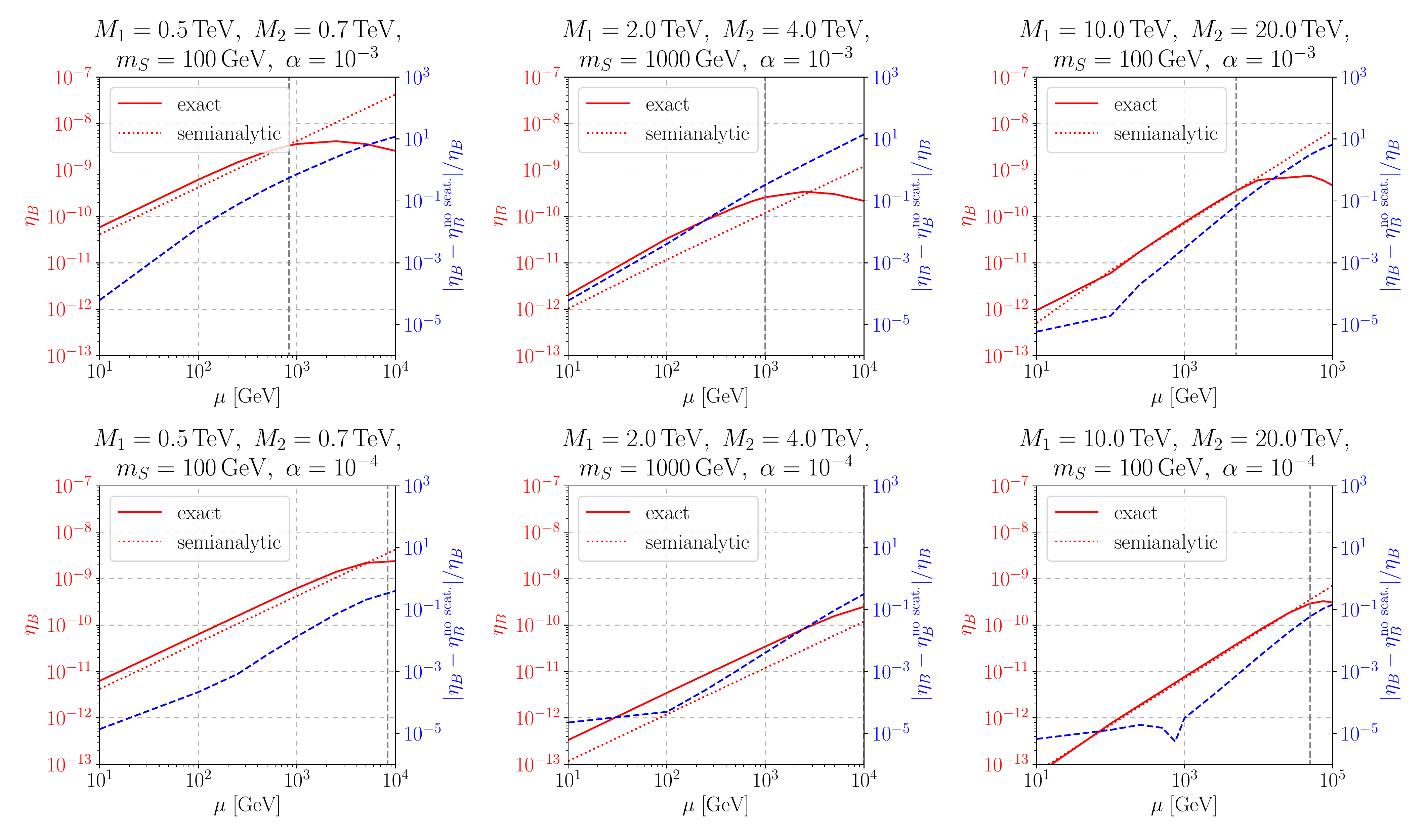}
\caption{\label{fig:muStarPlot}Comparison of the baryon asymmetry obtained via the solution of the full set of Boltzmann equations including scatterings (solid red line) and our estimate in Eq.~\eqref{eq:fit} (dashed red line) for different masses (columns) and couplings $\alpha_{ij} = 10^{-3}, 10^{-4}$ (rows) as indicated above each plot. We observe that, as long as $\mu<\mu_*$ [Eq.~\eqref{eq:scatteringEst}], the results agree well, and the true baryon asymmetry peaks for a value of $\mu$ close to $\mu_*$ (dashed vertical line). The blue curve indicates the ratio of the final baryon asymmetry obtained numerically by solving the Boltzmann equations with and without the scatterings. The scatterings are important when this ratio grows to be $\mathcal{O}(1)$, which in turn defines $\mu_*$. We checked that the approximation is valid for a larger range of parameters, but do not show this here explicitly.}
\end{figure*}


\subsection{Washout effects due to scattering processes}
\label{sec:Scattering}


The complete set of Boltzmann equations including different scattering terms was given in Ref.~\cite{Dall:2014nma}. The dominant washout effect is induced by the scatterings $N_iN_j\to HH$ mediated via the scalar singlet. While the asymmetry factors can be enhanced by increasing the trilinear coupling, $\mu$, this simultaneously enhances the scatterings, as can be seen from the scattering cross section~\cite{Alanne:2017sip}
\begin{equation}
	\label{eq:1}
	\sigma_{N_iN_j\rightarrow HH}=\frac{|(\alpha \alpha^{\dagger})_{ji}|^2\mu^2}{8\pi}\frac{s-(M_i+M_j)^2}{(s-m_S^2)^2
	    \sqrt{\delta_{M_i,M_j}}},
    \end{equation}
    where
    \begin{equation}
	\label{eq:2}
	\delta_{M_i,M_j}=(s-M_i^2-M_j^2)^2-4M_i^2M_j^2.
    \end{equation}
The scattering functions, $S_{N_iN_j\to HH}$, in Eqs.~\eqref{eq:BoltzmannEqs} can be written in terms of the scattering cross sections as~\cite{Dall:2014nma}
    \begin{equation}
	\label{eq:4}
	\begin{split}
	&S_{N_iN_j\rightarrow HH}\equiv\frac{M_i}{64\pi^2H(T=M_i)}\\
	&\qquad\times\int_{w_{\mathrm{min}}}^{\infty}\mathrm{d}w\sqrt{w}K_1(\sqrt{w})\hat{\sigma}_{N_iN_j\rightarrow HH}\left(\frac{w M_i^2}{z_i^2}\right),
	\end{split}
    \end{equation}
    where $w_{\mathrm{min}}=(M_i+M_j)^2$, and
    \begin{equation}
	\label{eq:}
	\hat{\sigma}_{N_iN_j\to HH}=\frac{1}{s}\,\delta_{M_i,M_j}\,\sigma_{N_iN_j\rightarrow HH}.
    \end{equation}
Comparing the analytical estimates and the exact result including these scattering terms, we find that the results agree well as long as
\begin{equation}\label{eq:scatteringEst}
    \mu < \mu_* \simeq M_1 \times \left(\frac{0.5}{\delta_M}\right) \left( \frac{10^{-3}}{\alpha} \right),
\end{equation}
where 
\begin{equation}
    \label{eq:deltaM}
    \delta_M \equiv \frac{M_2 - M_1}{M_1},
\end{equation} 
and we have taken all $\alpha_{ij}$ equal and denote them here and in the following collectively by $\alpha$.
Moreover, we observe that, since these scattering processes are relevant only to the washout, the regime $\mu > \mu_*$ will exhibit \emph{less} $B-L$ asymmetry than the regime for which Eq.~\eqref{eq:scatteringEst} holds. Thus, the baryon asymmetry is maximal for a value $\mu \sim \mu_*$, where the trilinear coupling is large enough to yield sufficient $CP$ asymmetry and yet not too large for scatterings to contribute significantly to the washout. This interplay is shown in Fig.~\ref{fig:muStarPlot}, where we compare in red (left scale) the final baryon asymmetries obtained via the semianalytical fit Eqs.~(\ref{eq:finalBasymmetry}, \ref{eq:fit}) (dotted line) and the exact numerical solution, including washout due to scattering (solid line). The dashed blue curve (right scale) shows the ratio of the numerical result with and without scatterings. When this ratio becomes $\mathcal{O}(1)$, scatterings tend to increase the washout and consequently the two red curves deviate. The scale $\mu_*$ is therefore chosen such that for $\mu<\mu_*$ scatterings are negligible and good agreement between numerics and Eqs.~(\ref{eq:finalBasymmetry}, \ref{eq:fit}) is achieved.


\begin{figure*}[t]
\centering
\includegraphics[width=0.48\textwidth]{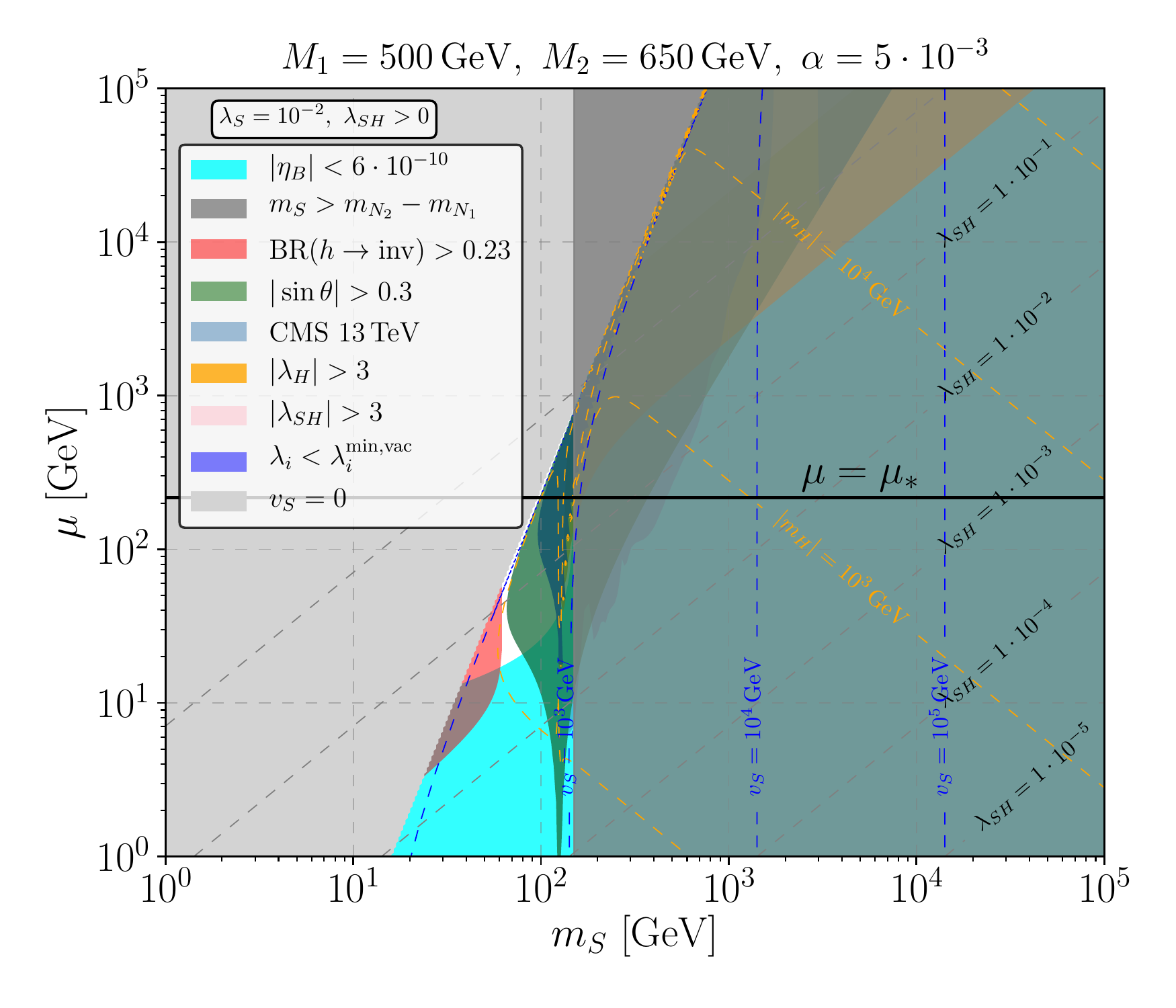}
\hfill
\includegraphics[width=0.48\textwidth]{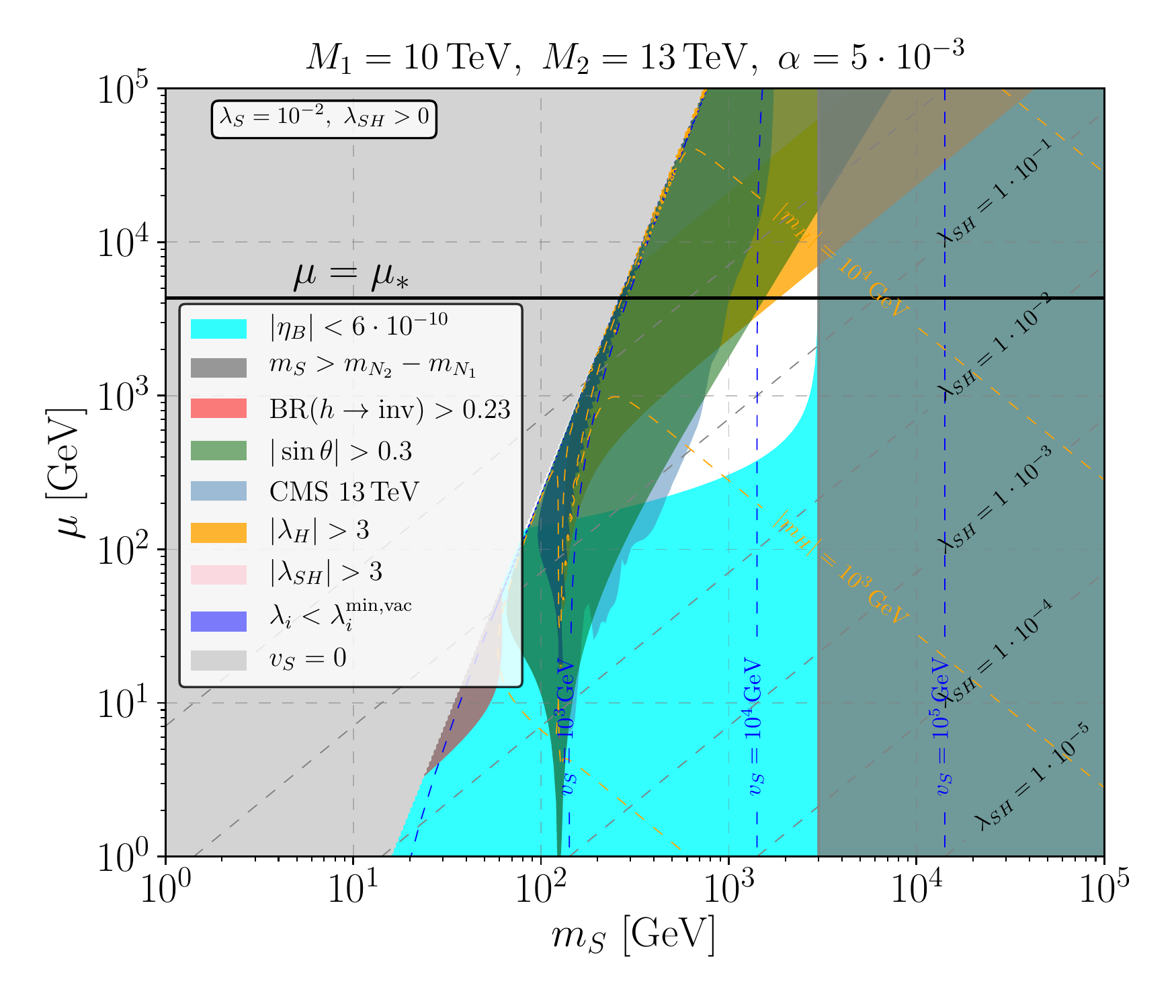}\\
\includegraphics[width=0.48\textwidth]{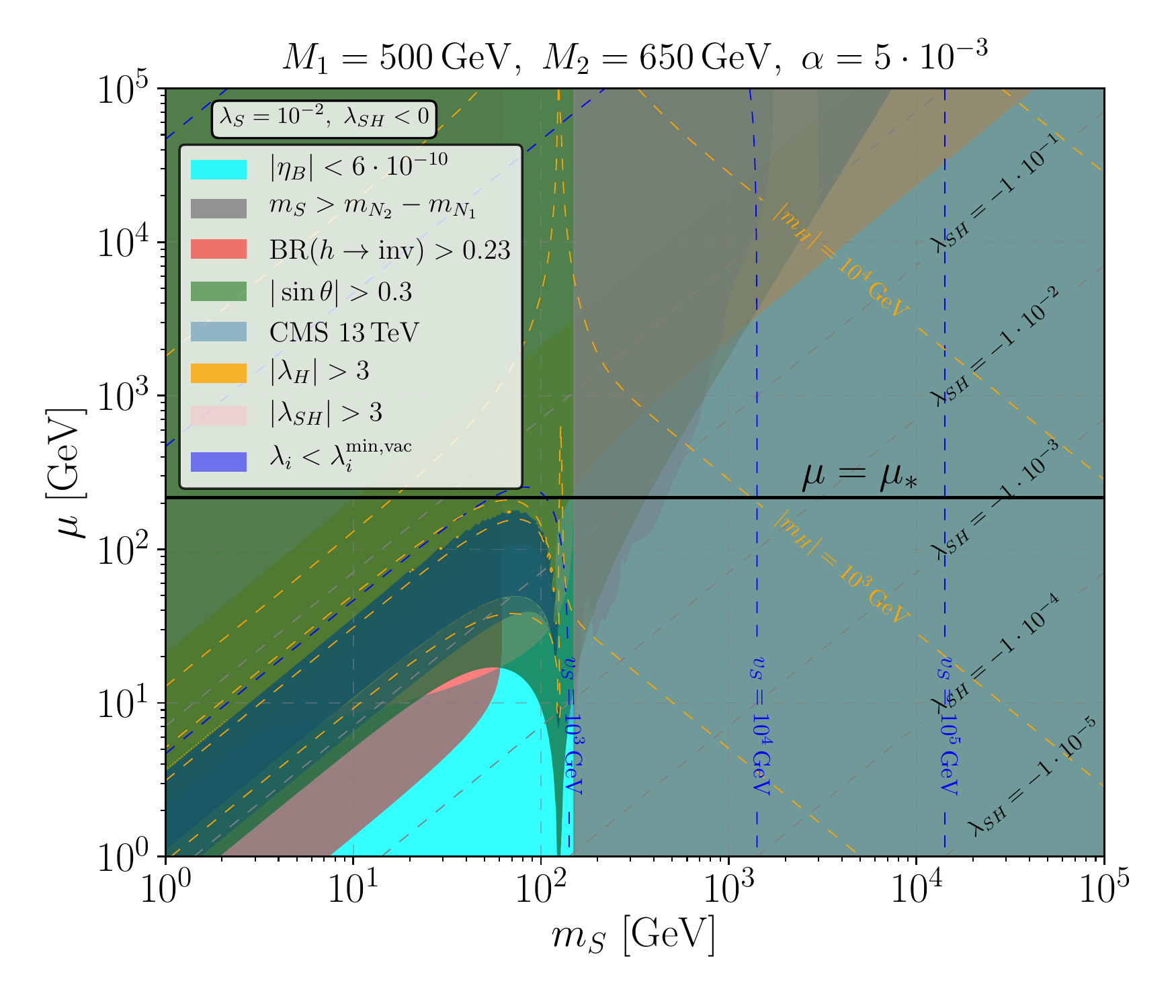}
\hfill
\includegraphics[width=0.48\textwidth]{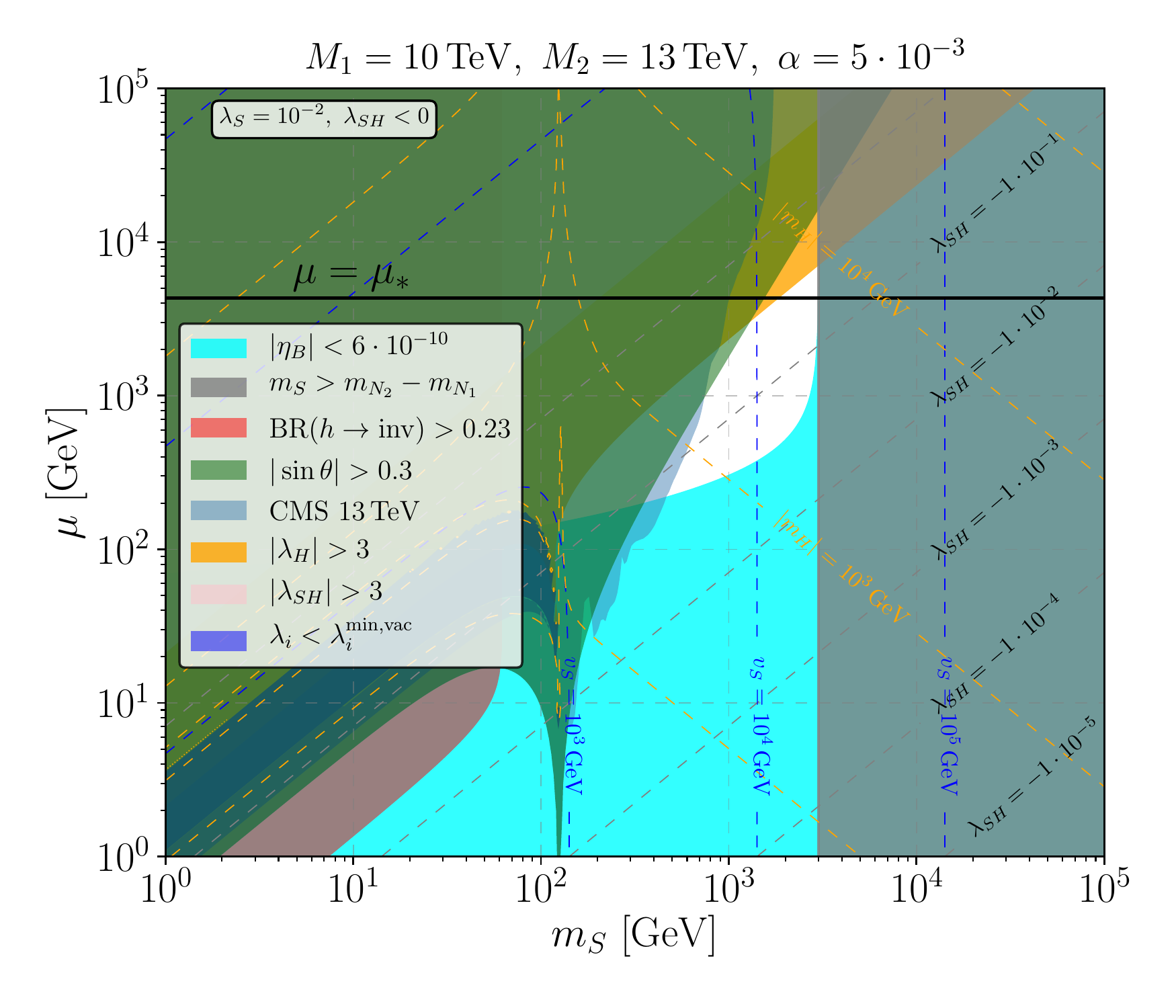}
\caption{\label{fig:ParamScan}Parameter scan obtained using our semianalytical approximation. We fixed the mass ratio, such that $\delta = 0.3$, which maximizes the resulting asymmetry. The CI angle is fixed such that $K_1 = 12$, $K_2 = 51$. Raising the mass scale from $500\GeV$ to $10\TeV$ opens a larger viable region, and at the same time raises the scale $\mu_*$. Thus, we find a lower bound of $M_1 \gtrsim 500\GeV$. }
\end{figure*}


\section{Viable parameter space for low-scale leptogenesis}
\label{sec:ScalarConst}


As a concrete and \textit{minimal} example, we consider
the $\mathbb{Z}_2$-symmetric scalar potential with a real singlet scalar:
\begin{equation}
\label{eq:Z2pot}
\begin{split}
V(H,\widetilde{S}) & = \widetilde{m}_H^2 |H|^2 +
\frac{1}{2}\,\widetilde{m}_S^2\,\widetilde{S}^2 + \lambda_H |H|^4\\
& + \frac{1}{2}\,\lambda_{SH}\,\widetilde{S}^2|H|^2 + 
\frac{1}{4}\,\lambda_S\,\widetilde{S}^4 \,.
\end{split}
\end{equation}
Here, we also adopted the notation of Sec.~\ref{sec:framework} that
the vacuum-dependent quantities are equipped with a tilde in the high-temperature phase,
where all VEVs vanish, and the quantities during leptogenesis are denoted without a tilde.
We first employ several theoretical and experimental constraints on parameter space: 
For the above perturbative treatment to be valid, the quartic couplings need to be small, naively ${\lambda_H, |\lambda_{SH}|<4\pi}$. We, however, adopt a more conservative bound $|\lambda_i| < 3$ in the light of perturbative unitarity bounds~\cite{Goodsell:2018tti}. Furthermore, to ensure that the scalar potential is bounded from below, we have to require that 
$\lambda_H,\lambda_S \geq 0$ and ${\lambda_{SH}>-2\sqrt{\lambda_H\lambda_S}}$, and the corresponding excluded region 
is indicated in Fig.~\ref{fig:ParamScan} by $\lambda_i < \lambda_i^{\text{min,vac}}$.


From the Higgs signal-strength measurements, the mixing angle between
the 125-GeV Higgs and the singlet is bounded by
$\sin\theta\lesssim 0.3$~\cite{Robens:2015gla,Ilnicka:2018def}. 
In addition, we include limits on the mixing angle from searches
for resonant scalar-singlet production at the LHC.
The limits we show correspond to those reported by the CMS
collaboration based on $36\,\textrm{fb}^{-1}$~of data
at $13\TeV$~\cite{Sirunyan:2018qlb,Buttazzo:2018qqp}. 
These searches give a stronger limit than the Higgs signal-strength measurements
for scalar-singlet masses $m_S \lesssim 800\GeV$.
Furthermore, for a very light scalar, $m_S<m_h/2$, the Higgs can decay into two singlets, and the invisible Higgs decays are currently bounded by
$\mathrm{BR}(h\to \mathrm{inv})\leq 0.23$~\cite{Aad:2015pla,Khachatryan:2016whc}. 
The width for the Higgs decay $h\to SS$ is given by
\begin{equation}
\Gamma_{h\rightarrow SS}=\frac{\lambda_{SH}^2v_{\rm ew}^2}{32\pi m_h} \sqrt{1-\frac{4m_S^2}{m_h^2}},
\label{eq:higgsdecaywidth}
\end{equation}
while the Higgs total
decay width to the visible SM channels is $\Gamma_h=4.07$ MeV for $m_h=125$ GeV~\cite{Dittmaier:2011ti}.


For the low-scale leptogenesis mechanism to work, the scalar singlet must acquire a nonzero VEV ($v_S > 0$) to induce the 
trilinear coupling $\mu S|H|^2$, and the condition that the decay $N_2\to N_1S$ has to be kinematically allowed constrains the singlet mass
during leptogenesis from above, ${m_S<M_1-M_2}$. In addition, we need the asymmetry to be generated before the EW phase transition, i.e. the phase transition in the singlet direction has to happen well above $T_\mathrm{sph}$. 
To avoid the false phase transition pattern, we fix as a benchmark value a small singlet self-coupling $\lambda_S=0.01$. This 
ensures that the thermal corrections to the singlet mass parameter are much smaller than to the Higgs mass (see, e.g.,~\cite{Alanne:2014bra}). For the resulting asymmetry, however, the numerical value of $\lambda_S$ is irrelevant.


With a complete understanding of the relevant bounds and an estimate for the maximal value of $\mu$, chosen such as to limit the washout by scattering processes, we can use the semianalytical estimate, Eq.~\eqref{eq:fit}, to efficiently scan the parameter space and identify viable regions. Including the above bounds, 
we obtain Fig.~\ref{fig:ParamScan}, where we fixed the mass splitting to $\delta_M = 0.3$. This modest degeneracy is accepted in order to achieve an optimal interplay between large $CP$ asymmetry and not too large washout. This allows us to identify the lowest RHN mass with successful leptogenesis; demanding less degeneracy implies larger RHN masses. The light blue region is excluded because there we obtain an insufficient baryon asymmetry ${|\eta_B| < 6\cdot 10^{-10}}$, whose sign can always be adjusted, e.g.,~via the sign of the CI rotation angle. The vertical gray line limits the scalar mass, because we need $m_S < M_2 - M_1$ for $\varepsilon_2^\mathrm{v/s} \neq 0$. For RHN masses $M_1 \lesssim 500\GeV$, this pushes the region with sufficient $CP$ violation beyond the bounds of the scalar singlet model, most importantly the light gray area where $v_S =0$, cf.~the upper left panel of Fig.~\ref{fig:ParamScan}. Therefore, $M_1 \approx 500 \,\textrm{GeV}$ is the lowest possible RHN mass which allows for successful baryogenesis, complies with all constraints that we considered and predicts an accompanying scalar DOF with $m_S \simeq 100\GeV$. Raising the RHN mass scale to, say, $M_1 = 10\TeV$, a larger viable region opens up (right panels of Fig.~\ref{fig:ParamScan}). Since at the same time $\mu_*$ grows, we find that the accompanying scalar can have a mass in the range $300\GeV \lesssim m_S \lesssim 3000\GeV$.
Remarkably enough, the viable parameter region in this case can be
probed by searches for resonant scalar-singlet production at current
and future colliders (see, e.g., the shaded
region labeled with ``CMS 13 TeV'').


\section{Conclusions and outlook}
\label{sec:conclusions}


Baryogenesis via leptogenesis is an appealing mechanism that connects the phenomenology of neutrino
oscillations at low energies to the generation of the BAU.
Standard thermal leptogenesis in the type-I seesaw model, however, requires very heavy right-handed neutrinos, 
making it hard to probe this scenario experimentally and resulting in large radiative corrections to the Higgs mass.
Currently, the most popular approaches for successful leptogenesis with lower RHN masses
rely either on resonant enhancement due to a highly degenerate RHN spectrum 
or on RHN flavor oscillations.

In this paper, we explored a particularly simple modification of the type-I seesaw framework that enhances $CP$ violation and the departure from thermal equilibrium such that the mass of the introduced RHNs can be accessed in Earth-based experiments. This modification relies only on two RHNs and one new singlet 
scalar and represents the low-energy effective theory of a large class of possible UV-complete models.
Coupling the singlet scalar, $S$, to the RHNs, $N_{1,2}$, opens up a new $N_2$ decay channel, $N_2 \rightarrow N_1 S$, 
leading to the above mentioned enhancement of $CP$ violation and departure from thermal equilibrium during the $N_2$ decays.
Together, these two effects increase the efficiency of leptogenesis,
such that the BAU can be successfully generated in $N_2$ decays
for RHN masses even below the TeV scale and without the need for
a highly degenerate RHN mass spectrum.

We derived a semianalytical fit function for the final baryon asymmetry
that allows for an efficient study of the parameter space, while reproducing the exact numerical result
with high precision. Using the semianalytical fit, we were able to determine the dependence of the final asymmetry on the choice of parameter values in the neutrino sector taking into account the 
the most recent low-energy data on neutrino oscillations.
We showed that the requirement of successful leptogenesis singles out
an interesting region in parameter space that can be probed in on-going
and future experiments.
In the present analysis, we were able to find viable scenarios 
with the lightest RHN mass down to roughly 500\,GeV, but
this bound should merely be regarded as a rough estimate.
We leave the refinement our analysis that would allow for
determining a more precise lower bound on $M_1$ for future work. 
Most importantly, Eq.~\eqref{eq:finalBasymmetry} should be generalized to take into account flavor effects, which we disregarded in the present analysis in order to focus on the new kinematic enhancement due to the singlet scalar. Furthermore, given that the lowest masses that we found lie around the EW scale, a more
detailed treatment of sphaleron freeze-out at temperatures around 
the EW phase transition should also be considered.

In addition, it would be interesting to embed our scenario into
concrete UV-complete models that would allow to predict some of our model
parameters such as the Yukawa matrix, $\alpha$, or the trilinear coupling, $\mu$.
However, all of these questions are beyond the scope of this work.
We conclude by emphasizing that the extension of the type-I seesaw model
by a real scalar singlet has exciting implications for low-scale leptogenesis
that call for further investigation.


\acknowledgments


The authors would like to thank Alexander Helmboldt for helpful discussions
on the real-scalar-singlet extension of the Higgs sector.
The authors are also grateful to Filippo Sala for discussions on the CMS searches
for resonant singlet production at the LHC as well as for making available
to us some of the numerical data presented in Ref.~\cite{Buttazzo:2018qqp}.
M.\,P.\ acknowledges funding from the International Max Planck Research School
for Precision Tests of Fundamental Symmetries (IMPRS-PTFS).
T.\,H.\ and M.\,P.\ are enrolled at the University of Heidelberg.


\onecolumngrid
\appendix


\section{Loop functions}
\label{app:formulas}


The loop functions relevant for the $CP$-asymmetry parameters 
are given by~\cite{Dall:2014nma}
\begin{subequations}
\begin{gather}
\mathcal{F}^{(\rm{v})}_{ij,\, R} = \sqrt{r_{ji}}\,\ln\left[\frac{(1+\eta_i)(1-r_{ji})- (1-\eta_i)(\sigma_i+\sqrt{\delta_{ji}})}{(1+\eta_i)(1-r_{ji})- (1-\eta_i)(\sigma_i-\sqrt{\delta_{ji}})}\right],\\
\mathcal{F}^{(\rm{v})}_{ij,\, L} = -\sqrt{\delta_{ji}} + \frac{r_{ji}- \eta_i}{1-\eta_i} \ln\left[\frac{(1+\eta_i)(1-r_{ji})- (1-\eta_i)(\sigma_i+\sqrt{\delta_{ji}})}{(1+\eta_i)(1-r_{ji})- (1-\eta_i)(\sigma_i-\sqrt{\delta_{ji}})}\right],\\
    \mathcal{F}^{(\rm{s})}_{ijk,\, RR} = \frac{\sqrt{r_{ji}}\sqrt{r_{ki}} \sqrt{\delta_{ji}}}{1-r_{ji}},
    \hspace{2.5cm}\mathcal{F}^{(\rm{s})}_{ijk,\, RL} = \frac{1}{2} \frac{\sqrt{r_{ki}} \sqrt{\delta_{ji}} (1+ r_{ji} - \sigma_i)}{1-r_{ji}}\\
    \mathcal{F}^{(\rm{s})}_{ijk,\, LL} = \frac{\sqrt{r_{ji}} \sqrt{\delta_{ji}}}{1-r_{ji}},
    \hspace{3cm}\mathcal{F}^{(\rm{s})}_{ijk,\, LR} = \frac{1}{2} \frac{\sqrt{\delta_{ji}} (1+ r_{ji} - \sigma_i)}{1-r_{ji}}\,,
\end{gather}
\end{subequations}
where $r_{ij} \equiv M_i^2 / M_j^2$, $\sigma_i \equiv m_S^2 / M_i^2$, $\eta_i \equiv m_h^2 / M_i^2$ and $\delta_{ij} \equiv (1-r_{ij} -\sigma_j)^2 - 4\, r_{ij} \sigma_j$. Note that during leptogenesis (before the EW phase transition), the zero-temperature Higgs boson mass vanishes, $m_h = 0$. This is not necessarily the case if a scalar doublet other than the SM Higgs is involved in leptogenesis, like for example in the scotogenic model.


\twocolumngrid
\bibliography{literature}
\onecolumngrid


\end{document}